\documentclass[manuscript]{aastex63}

\accepted{June 24, 2020}
\submitjournal{ApJ}

\shorttitle{Parker-Jeans instability with cosmic rays}
\shortauthors{Kuwabara \& Ko}

\begin{document}

\title{Dynamics of the Parker-Jeans instability of gaseous disks including the effect of cosmic rays}

\author{Takuhito Kuwabara}
\affiliation{Computational Science and Engineering Division II, \\
AdvanceSoft Corporation, 4-3, Kanda Surugadai, Chiyoda-ku, Tokyo 101-0062, Japan}

\author{Chung Ming Ko}
\affiliation{Institute of Astronomy, Department of Physics and Center for Complex Systems, \\
National Central University, Zhongli Dist., Taoyuan City, Taiwan 320, Republic of China}



\begin{abstract}
The effect of cosmic rays on the Parker-Jeans instability in magnetized self-gravitating
gaseous disks is studied by three-dimensional magnetohydrodynamic (MHD) simulations with
cosmic rays taken as a massless fluid with notable pressure. Cosmic ray diffusion along the
magnetic field is taken into account in the simulation.
The initial configuration of the disk is a magnetized cold gas slab sandwiched between
hot corona (on top and bottom).
We show that cosmic rays play an important role in the formation of filaments or clumps in some parameter regimes.
In a certain range of the thickness of the gas slab, the cosmic ray diffusion coefficient plays
a decisive role in determining whether the filaments lie along or perpendicular to the magnetic field.
We also consider the effect of rotation on the system.

\end{abstract}



\section{Introduction} \label{sec:intro}
In our Galaxy, the interstellar medium (ISM) comprises different components, such as different phases of gas, magnetic field and cosmic rays.
All these components have similar energy density \citep[e.g.,][]{Parker_1969, Ginzburg_1976, Ferri_re_2001, Cox_2005}.
It is understandable that cosmic ray is dynamically important in the structure and evolution of ISM,
yet not many studied were devoted to the role played by cosmic ray on ISM.
Nevertheless, in the past couple of decades efforts have been made on the influence of cosmic ray on instabilities
(say, Parker instability, magneto-rotational instability)
\citep[e.g.,][]{Parker_1966, Kuznetsov_1983, Hanasz_1997, Hanasz_Lesch_1997, Ryu_2003,
Kuwabara_2004, Kuwabara_2006, Hanasz_2009, Ko_2009, Lo_2011, Kuwabara_2015, Heintz_2018},
and on cosmic ray modified structures and outflows
\citep[e.g.,][]{Ghosh_1983, Ko_1987, Ko_1991, Ko_1991a, Ko_1991b, Breitschwerdt_1991, Breitschwerdt_1993,
Everett_2008, Yang_2012, Girichidis_2016, Recchia_2016, Wiener_2017, Pfrommer_2017, Farber_2018}.
Generally speaking, cosmic rays often enhance instabilities.
They help drive galactic winds (but may hinder stellar winds)
The diffusion of cosmic ray can affect the growth rate of instability. 
For instance, the growth rate of the Parker instability becomes larger if the diffusion coefficient of cosmic ray is larger \citep[][]{Kuwabara_2004}.
Moreover, we note that cosmic ray diffusion may have some subtle effect on the dynamics of the system.
The present work will illustrate one example.

An important subject in molecular cloud (MC) and star formation research is the relation between magnetic field and molecular clouds 
\citep[e.g., see the review by][]{Crutcher_2012}. 
The orientation between magnetic field and cloud filaments or cores reveals the dynamics of cloud collapse. 
\citet[][]{Tassis_2009} derived the intrinsic shapes and magnetic field orientations of 24 MCs by
statistical analysis using dust emission and polarization data from the Hertz polarimeter.
They showed that the best-fitting intrinsic magnetic field orientation is close to the direction of the minor axis of the oblate disks.
\citet[][]{Li_2013} made use of near-infrared dust extinction maps and optical stellar polarimetry  to compare the orientations between 
13 filamentary clouds in the Gould Belt and their local jntercloud media magnetic fields.
They obtained a bimodal distribution in which the clouds tend to be either parallel or perpendicular to the mean direction of the magnetic field.
\citet[][]{Soler_2013} studied the relative orientation of the magnetic field with respect to the density structures by
synthetic observations of the simulated turbulent molecular clouds. 
They adopted the method Histogram of Relative Orientations(HRO),
which utilized the gradient to characterize the directionality of column density structures on multiple scales.
They concluded that in most cases the orientation of the magnetic field is parallel to the density structure.
However, in strongly magnetized cases, the orientation changes from parallel to perpendicular where the density is higher than a critical density.
\citet[][]{Planck_2016} evaluated the relative orientation of the magnetic field inferred from the Planck polarization observations at 353GHz 
with respect to the gas column density structures for 10 nearby Gould belt MCs by means of HRO.
They found that the relative orientation changes from parallel to perpendicular with increasing column density.

The bulk of cosmic rays in ISM is low energy (below a few hundred MeV).
As they travel through ISM, they lose energy via ionization (and through damping of waves they excited).
Increase in ionization rate can heat up gas and hinder diffusion of magnetic field, thus affects star formation processes
\citep[e.g.,][]{Fatuzzo_2006, Yusef_Zadeh_2007, Glassgold_2012, Bertram_2015}.
We are interested in the dynamical influence of cosmic rays on star formation, in particular,
the formation and development of clouds.

\citet{Chou_2000} studied the dynamics of the Parker-Jeans instability by linear stability analysis and MHD simulation.
They showed the process of the interstellar gas aggregation to molecular clouds.
\citet{Kuwabara_2006} added cosmic rays into the system and showed, by linear stability analysis, that
the self-gravitating gaseous disks is less unstable if cosmic ray pressure is larger,
and more unstable if the cosmic ray diffusion coefficient is larger.
However, the nonlinear development of the system has not been investigated yet.
In view of recent progresses in numerical techniques in MHD simulation with cosmic ray, we would like to revisit the problem of
Parker-Jeans instability of a disk until the nonlinear stage.
In the case of no cosmic ray diffusion, the set of MHD equations with cosmic rays can be written in fully conservation form
(and cosmic ray can be expressed as a polytropic gas, $P_c\propto \rho^{\gamma_c}$) \citep[][]{Kudoh_2016}.
The set of equations can then be simulated more precisely, e.g., in dealing with shock problems.
The treatment of the cosmic ray diffusion, which is the parabolic term in the cosmic ray energy equation,
has more restrictive time step constraints than that in the system without cosmic ray diffusion for explicit methods.
Implicit methods can overcome such restriction, but they involve inverting large matrix which is computationally expensive.
Super-time-stepping methods \citep[e.g.,][]{Alexiades_1996} is a tradeoff between explicit and implicit methods in this regard.
These methods can be viewed as an explicit Runge-Kutta method with several internal stages by using the recursion relations associated
with Chebyshev Polynomials.
\citet{Meyer_2012} presented a better stability super-time-stepping method which is a multi-stage Runge-Kutta method
based on the recursion sequence for Legendre polynomials instead of Chebyshev Polynomials.
Usually, super-time-stepping methods are used in solving the heat conduction problem.
On the other hand, it is possible to be applied in solving the cosmic ray diffusion problem.
In this paper, we applied this method to solve the anisotropic cosmic ray diffusion problem,
and study the effect of cosmic rays on the Parker-Jeans instability by MHD simulations.
Linear stability analysis is supplemented for comparison.

This paper is organized as follows.
In Section~\ref{sec:models} we present the governing equations of the self-gravitating disk
and the initial equilibrium model, the two temperature layered disk.
In Section~\ref{sec:floats} the results of MHD simulations are presented.
Section~\ref{sec:summary} provides a summary and discussion.

\section{Models} \label{sec:models}
\subsection{Two-fluid self-gravitating disk} \label{subsec:two-fluid}

We adopt a two-fluid MHD system. One fluid is the common magnetized thermal plasma and the other one is cosmic ray.
Cosmic ray is considered as a massless fluid with notable energy density (or pressure).
Cosmic ray is coupled to the plasma via magnetic fluctuations, resulting in cosmic ray advection and diffusion in the plasma.
The energy exchange between the plasma and cosmic ray is facilitated by the work done of cosmic ray pressure gradient.
In a rotating frame, the system is governed by the total mass, momentum and energy equation of the system,
\begin{eqnarray}\label{mass-total}
  {\partial\rho\over\partial t}+{\bf\nabla}\cdot(\rho{\bf V})=0\,,
\end{eqnarray}
\begin{eqnarray}\label{momentum-total}
 {\!\partial\over\partial t}\left(\rho{\bf V}\right)+
  {\bf\nabla}\cdot\left[\rho{\bf V}{\bf V}+\left(P_{\rm g}+P_{\rm c}+{B^2\over 2\mu_0}\right){\bf I}-
  {{\bf B}{\bf B}\over\mu_0}\right] \nonumber \\
  =-\rho\left[\nabla\psi-{\bf g_{\rm ext}}+2{\bf\Omega}\times{\bf V}
   +{\bf\Omega}\times\left({\bf \Omega}\times{\bf r} \right) \right],
\end{eqnarray}
\begin{eqnarray}\label{energy-total}
{\partial\over\partial t} \left( E+E_{\rm c}+{B^2\over 2\mu_0}\right)
+\nabla\cdot\left[
\left(E+E_{\rm c}+P_{\rm g}+P_{\rm c}\right){\bf V}
-\frac{\left({\bf V}\times{\bf B}\right) \times{\bf B}}{\mu_0}
\right] \nonumber \\
=\nabla\cdot\left(\kappa_{\|}{\bf b}{\bf b}\cdot \nabla E_{\rm c}\right)
-\rho{\bf V}\cdot\left[\nabla\psi-{\bf g_{\rm ext}}+{\bf\Omega}\times\left({\bf \Omega}\times{\bf r} \right) \right],
\end{eqnarray}
supplemented by the cosmic ray energy equation, the induction equation for magnetic field and the Poisson equation for self-gravity,
\begin{eqnarray}\label{cr-energy-1}
\frac{\partial E_{\rm c}}{\partial t}+\nabla\cdot\left[\left(E_{\rm c}+P_{\rm c}\right){\bf V}\right]
={\bf V}\cdot\nabla P_{\rm c}
+\nabla\cdot\left(\kappa_{\|}{\bf b}{\bf b}\cdot \nabla E_{\rm c}\right),
\end{eqnarray}
\begin{eqnarray}\label{mag-energy}
  {\partial{\bf B}\over\partial t}-{\bf\nabla}\times({\bf V}\times{\bf B})=0\,,
\end{eqnarray}
\begin{eqnarray}\label{poisson}
\nabla^2\psi=4\pi G\rho\,.
\end{eqnarray}
where $E=E_{\rm k}+E_{\rm g}=\rho V^2/2+P_{\rm g}/(\gamma_{\rm g}-1)$ is the sum of kinetic and thermal energy density of the plasma;
$\rho$, $\bf V$, $P_{\rm g}$, and $\gamma_{\rm g}$ are the plasma density, flow velocity, thermal pressure, and polytropic index;
$E_{\rm c}=P_{\rm c}/(\gamma_{\rm c}-1)$, $P_{\rm c}$ and $\gamma_{\rm c}$ are the energy density, pressure and the polytropic index for cosmic ray;
$\psi$ and $\bf g_{\rm ext}$ are the gravitational potential for self-gravity and the external gravitational acceleration;
$\bf \Omega$ is the rotational angular frequency;
$\bf B$, $\bf b={\bf B}/|{\bf B}|$ are the magnetic field and the unit vector in the direction of magnetic field;
$\kappa_\|$ is the cosmic ray diffusion coefficient along the magnetic field; and $\bf I$ is the unit tensor.

\subsection{Equilibrium model} \label{subsec:equil}
We set forth to study a local slab portion of a rotating, self-gravitating disk.
We adopt a local Cartesian coordinate system $(x,y,z)$ such that ${\bf e}_x={\bf e}_\phi$, ${\bf e}_y=-{\bf e}_r$, and ${\bf e}_z={\bf e}_z$,
where $(r,\phi,z)$ is the cylindrical coordinate system of the disk.
We set up a simple hydrostatic equilibrium model as the initial background configuration for the simulation.
Assume the centrifugal force is balanced by the gravitational force in the horizontal direction, and all other quantities depend on $z$ only.
In addition, assume the magnetic field is lying horizontally (and there is no cross field line diffusion of cosmic ray).
Then with ${\bf V}=0$, Equations~(\ref{mass-total}), (\ref{energy-total}), (\ref{cr-energy-1}) \& (\ref{mag-energy}) are satisfied automatically.
There are only two equations left.
The Poisson equation (Equation~(\ref{poisson})),
\begin{eqnarray}\label{poisson2}
\frac{d^2 \psi}{d z^2}=4\pi G\rho\,,
\end{eqnarray}
and the momentum equation (Equation~(\ref{momentum-total})), which becomes the magneto-hydrostatic equation ($P_{\rm B}=B^2/2\mu_0$),
\begin{eqnarray}\label{hydrostatic}
\frac{1}{\rho}\frac{d P_{\rm t}}{dz}+\frac{d\psi}{dz}
=\frac{1}{\rho}\frac{d}{dz}\left(P_{\rm g}+P_{\rm B}+P_{\rm c}\right)+\frac{d\psi}{dz}
=g_{\rm ext}=0\,,
\end{eqnarray}
where $P_{\rm t}$ is the total pressure and $g_{\rm ext}$ is the external gravity due to other sources,
for example, the stellar disk in the case of the Galactic disk.
Eliminating $\psi$ from Equations~(\ref{poisson2}) \& (\ref{hydrostatic}), we obtain
\begin{eqnarray}\label{background}
\frac{d}{dz}\left(\frac{1}{\rho}\frac{dP_{\rm t}}{dz}\right)+4\pi G\rho
=\frac{d^2 h_{\rm t}}{dz^2}+4\pi G\rho
=\frac{dg_{\rm ext}}{dz}\,,
\end{eqnarray}
where $h_{\rm t}=\int dP_{\rm t}/\rho$ can be called the total enthalpy.
This equation was derived by \citet[][]{Chou_2000} and extended for taking into the effect of CRs in this work.
If the equation of state $P_{\rm t}=P_{\rm t}(\rho)$ is given, then Equation~(\ref{background}) can be solved.
We note that $P_{\rm g}$, $P_{\rm B}$ and $P_{\rm c}$ are not constrained by their energy equations (as they are satisfied automatically).
Thus, for simplicity, we take $P_{\rm B}=\alpha P_{\rm g}$ and $P_{\rm c}=\beta P_{\rm g}$, and adopt a polytropic equation of state
for the gas $P_{\rm g}\propto \rho^{\gamma_{\rm g}}$.
We then have $h_{\rm t}=C_{\rm s}^2(1+\alpha+\beta)/(\gamma_{\rm g}-1)$,
where $C_{\rm s}^2=\gamma_{\rm g} P_{\rm g}/\rho$ is the gas sound speed.

Furthermore, assume that the equilibrium gas layer is sandwiched between high-temperature gas layers given by,
\begin{eqnarray}
T(z)=0.5\times\left[T_{\rm cor}+T_0+(T_{\rm cor}-T_0)\times
{\rm tanh}\left(\frac{|z|-z_{\rm cor}}{\Delta z}\right) \right],
\end{eqnarray}
where $z_{\rm cor}$ and $\Delta z$ are the half thickness of the cold gaseous disk and the width of transition region
between the cold gas and hot gas layer,
$T_0$ and $T_{\rm cor}$ are the temperatures of the cold gas layer and hot gas sandwiching the cold gas layer, respectively.
In \citet{Kuwabara_2006}, $T_{\rm cor}$ was set as infinity for the linear stability analysis.
The initial equilibrium condition is obtained by solving Equation~(\ref{background}) numerically using Runge-Kutta method.

In the following MHD simulations, it is set as finite value \citep[][]{Shibata_1989}.
The scale height of the density is defined as
$H=C_{{\rm s}0}\sqrt{(1+\alpha+\beta)/(2\pi G\rho_{0}\gamma_{\rm g})}$ where the subscript $0$ denotes the value at the mid-plane.
Quantities are normalized to the following density, velocity and length, $\rho_0$, $C_{s0}$, and $H_0=C_{{\rm s}0}/\sqrt{2\pi G\rho_0\gamma_{\rm g}}$.
As fiducial values, we pick
$\rho_{0}=1.67\times10^{-19}\,{\rm kg~m^{-3}}$, $C_{{\rm s}0}=5\,{\rm km~s^{-1}}$,
$H_{0}=20\,{\rm pc}$, $\gamma_{\rm g}=1.05$ ,$\gamma_{\rm c}=4/3$
and the unit of time is $H_0/C_{s0}\sim 4~{\rm Myr}$.
The cosmic ray diffusion coefficient is estimated to be $3\times 10^{23}\,{\rm m^2~s^{-1}}$
\citep[e.g.,][]{Berezinskii_1990},
and the normalized diffusion coefficient $\kappa_{\|}$ is 100.
Here, we neglect the external gravity because it does not have a significant influence
on the dynamics of the system when the ratio of the external gravity to the self gravity
is less than one \citep[e.g.][]{Chou_2000}.

\section{Three-dimensional CR-MHD simulation with self-gravity} \label{sec:floats}
\subsection{Numerical procedure} \label{subsec:procedure}
We solve the MHD equations supplemented by the cosmic ray energy equation and the Poisson equation for self-gravity by numerical simulation.
We adopt a Harten-Lax-van Leer Discontinuities (HLLD) method \citep[][]{Miyoshi_2005} for the advection part of the numerical solver.
The self-gravity part (Poisson equation) is solved by the finite difference method,
and the large matrix inversion by the biconjugate gradients stabilized (BICGStab) method.
We apply a super-time-stepping scheme called second-order accurate $s$-stage Runge-Kutta Legendre scheme (RKL2) \citep[e.g.][]{Meyer_2012}
to solve the cosmic ray diffusion part in the cosmic ray energy equation,
which is the parabolic mathematically and is known to be computationally expensive.
Usually, this part is solved by implicit scheme to prevent from the restrictive time step constraints.
However, we already have applied the implicit method to solve the Poisson equation for self-gravity,
and it will be too computationally costly to apply again the implicit method for the cosmic ray diffusion part.
Thus, we select the super-time-stepping scheme as the computational cost of super-time-stepping is somewhere in between implicit and explicit methods.

\begin{figure}
\plotone{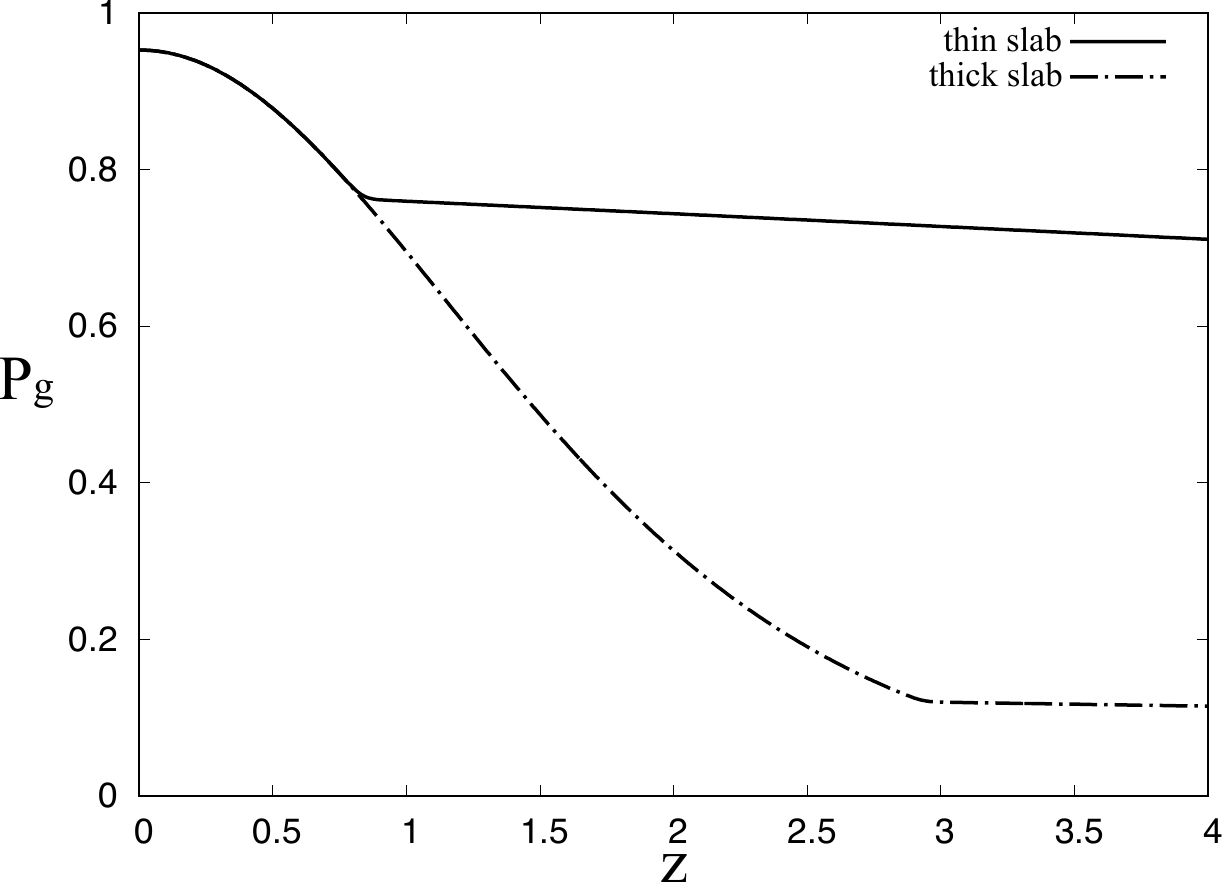}
\caption{Initial thermal pressure distribution in $z$-direction
         for the thin slab case and the thick slab case.}
\label{fig:zc_pr}
\end{figure}

We calculate a slab portion of rotating or non-rotating, self-gravitating disk in Cartesian coordinate.
The models that we studied are listed in Table~\ref{tab:models}.
In all models, the initial ratios of magnetic field pressure to gas pressure $\alpha$
and cosmic ray pressure to gas pressure $\beta$ are set to one.
The initial magnetic field of all model is ${\bf B}(z){\bf e}_x$ (i.e., in the azimuthal direction ${\bf e}_\phi$ of the disk).
The thickness of the slab is thin ($z_{\rm cor}=0.9$) for model~1, model~2, model~3, model~5, model~6, and ($z_{\rm cor}=0.6$) for model~7,
and thick ($z_{\rm cor}=3.0$) for model~4 and model~8.
The cosmic ray diffusion coefficient is high ($\kappa_{\|}=100.0$) for model~1 and model~4,
middle ($\kappa_{\|}=10.0$) for model~7, and low ($\kappa_{\|}=1.65$) for model~2, model~5 and model~8,
and no-diffusion for model~3 and model~6.
The rotation effect is applied only in model~5.
The $x$-direction, $y$-direction, and $z$-direction correspond to the azimuthal direction, the inward radial direction,
and the rotation axis of the disk.
The size of the simulation box in $x$-, and $y$-direction for each model
is decided from the wavelength of the maximum growth rate
given by the linear analysis and is shown as $\lambda_{x\,{\rm max}}$ and $\lambda_{y\,{\rm max}}$ in Table~\ref{tab:models}.
Figure~\ref{fig:zc_pr} shows the initial gas pressure distribution
for the thin slab case and the thick slab case.
The number of grid point in each direction is
$\left( n_{\rm x}\times n_{\rm y}\times n_{\rm z}\right)=\left( 100\times 100\times 200\right)$.
We assume periodic boundaries for $x=x_{\rm min}$ and $x=x_{\rm max}$, $y=y_{\rm min}$ and $y=y_{\rm max}$,
and free boundary for $z=z_{\rm min}$ and $z=z_{\rm max}$.

\subsection{Numerical results} \label{subsec:results}
In this subsection, we show the results of CR-MHD simulation on the formation of self-gravitating clouds
by imposing random perturbation to the initial equilibrium state described in Section~\ref{subsec:equil}.
The imposed perturbation is an velocity perturbation in the horizontal plane $\delta V_x$, $\delta V_y$
whose amplitude is distributed randomly between
$-0.05\le \delta V_x,\, \delta V_y\le 0.05$ \citep{Chou_2000}.
Figures~\ref{fig:model1}--\ref{fig:model8} show the results of model~1 to model~8 consecutively,
in which the time for model~1 is $t=17.5$, model~2 is $t=23.5$, model~3 is $t=23.5$,
model~4 is $t=11.0$, model~5 is $t=32.0$, model~6 is $t=19.0$, model~7 is $t=22.0$, and model~8 is $t=15.5$,
where the unit of time is about 4 Myr.
The left panel in each figure shows the normalized density distribution and the magnetic field lines.
The isosurface shows the normalized density at value equals to $1.7$,
and the lines show the magnetic field lines.
The right panel in each figure shows the normalized cosmic ray and thermal gas pressure distribution on the plane $z=0.0$.
Cosmic ray pressure is in color-scale, and thermal gas pressure in contours
(the range of contours is $1.0 \le P_{\rm g} \le 3.0$ with interval 0.5).

Initially, the cold gas is distributed uniformly in $x$- and $y$-direction.
As time proceeds, Parker-Jeans instability causes the gas to coalesce,
but it develops into different structures in different models.
In model~1, model~4, model~5, model~6 and model~8, filamentary structures are formed with their long-axis perpendicular to the magnetic field.
In model~2, the filaments break up into clumps.
In model~3 (the one without cosmic ray diffusion), the filamentary structures with long-axis parallel to the magnetic field are formed.
These results show that the cold gas coalesces or collapses to form filamentary clouds.
Depending on the cosmic ray diffusion coefficient and rotation, the filaments may align with or perpendicular to the magnetic field.
When the diffusion coefficient is large/small (model~1/model~3), long filaments are form perpendicular/parallel to the magnetic field,
and if the diffusion coefficient is somewhere in between (model~2), the filaments may turn into clumps with weak directionality.
When compare with thin slab cases (model~1, model~2, model~3, model~6 or model~7)
and thick slab case with small cosmic ray diffusion coefficient (model~8),
the deformation of magnetic field lines is larger in the thick slab case with large cosmic ray diffusion coefficient (model~4) and in
the case with rotation (model~5).
Moreover, in the thick slab case the range of thermal pressure variation between the mid-plane ($z=0$) and the half thickness ($z=z_{\rm cor}$)
is wider than the thin slab cases.
The effect of magnetic buoyancy is larger in the thick slab case with large cosmic ray diffusion coefficient.

The distribution of the cosmic ray pressure matches well with thermal gas pressure in model~2, model~3, model~5,
model~7 and model~8,
while they are almost uncorrelated in model~1 and model~4.
The cosmic ray diffusion coefficient of model~1 and model~4 is large, so that the cosmic ray pressure is nearly uniform.
As a result, the contribution of cosmic ray on cold gas coalescent is weak.
On the other hand, when the diffusion coefficient is small (model~2, model~3, model~5, model~7 and model~8),
cosmic ray pressure gradient is more significant
and the distribution of gas is strongly affected
\citep[][]{Kuwabara_2004}.

Rotation is considered in model~5.
Using the linear stability analysis method in \citet{Kuwabara_2006}, we work out how the maximum linear growth rate of model~5
depends on different angular velocity $\Omega$
(i.e., parameters other than $\Omega$ are $\alpha=1.0$, $\beta=1.0$, $\kappa_\|=1.65$ and $z_{\rm cor}=0.9$).
Figure~\ref{fig:sigma_omega} shows the dependence of the maximum growth rate on $\Omega$.
In the figure, $\sigma_{x\,{\rm max}}$ is the maximum growth rate in $x$-direction,
i.e., the maximum growth rate of perturbations which does not depend on $y$; and
similar definition applies to $\sigma_{y\,{\rm max}}$.
$\sigma_{y\,{\rm max}}$ decreases as $\Omega$ increases and becomes zero for $\Omega\ge 0.27$.
Hence, for large enough $\Omega$, perturbations in $x$-direction outgrow those in $y$-direction.
We set $\Omega=0.3$ in model~5 such that $\sigma_{y\,{\rm max}}=0.0$.
Figure~\ref{fig:model5} indicates that the gas forms long filaments perpendicular to the magnetic field,
i.e., perturbation variations grow predominately in $x$-direction.
This is consistent with the prediction of the linear stability analysis.

Figure~\ref{fig:time_den} shows the time evolution of the perturbed gas density
at the position where the density has its maximum value at the end of the simulation.
The growth is fastest in model~4, the thick slab case with large cosmic ray diffusion coefficient.
For thin slab cases, the growth of model~1 is faster than model~2 which in turns faster than model~3.
We deduce that the smaller the cosmic ray diffusion coefficient, the smaller the growth rate
\citep[][]{Kuwabara_2004}.
The smallest growth rate is model~5, in which the instability is suppressed by the effect of rotation.
All models evolve linearly at first and shift to nonlinear stage later,
and the gas cloud collapses (density tends to large values) eventually.

\section{Summary and discussion} \label{sec:summary}
We succeeded in carrying out analysis of the evolution of a self-gravitating two temperature
layered gas slab by MHD simulation and cosmic rays.
The gas slab is susceptible to Parker and Jeans instabilities.
Cosmic rays play an interesting dynamical role, in particular, when diffusion of cosmic ray
is take into account. We considered diffusion along the magnetic field only.

\begin{figure}
\plotone{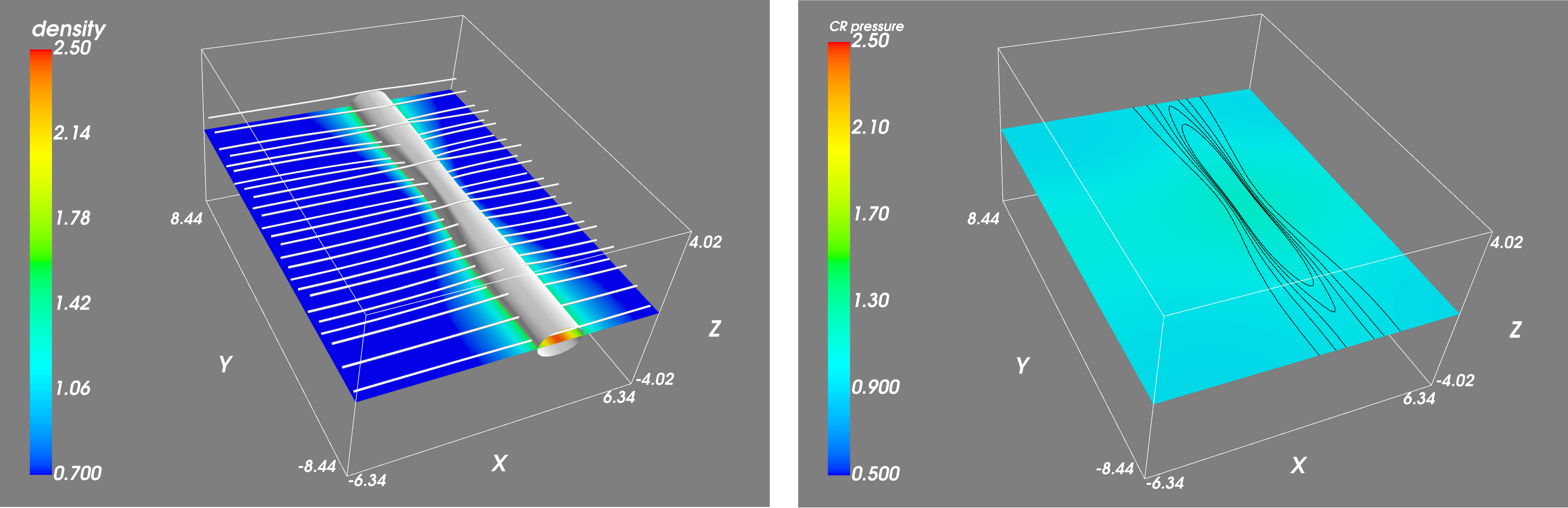}
\caption{Simulation result of model~1
         ($\alpha=1.0$, $\beta=1.0$, $\kappa_\|=100.0$, $z_{\rm cor}=0.9$, $\Omega=0.0$) at $t\sim70\,{\rm Myr}$.
         {\it Left}: Distribution of density (isosurface and color scale) and magnetic field lines.
         {\it Right}: Distribution of cosmic ray pressure (color scale) and gas pressure (contours)
         on the $z=0.0$ plane.}
\label{fig:model1}
\end{figure}

\begin{figure}
\plotone{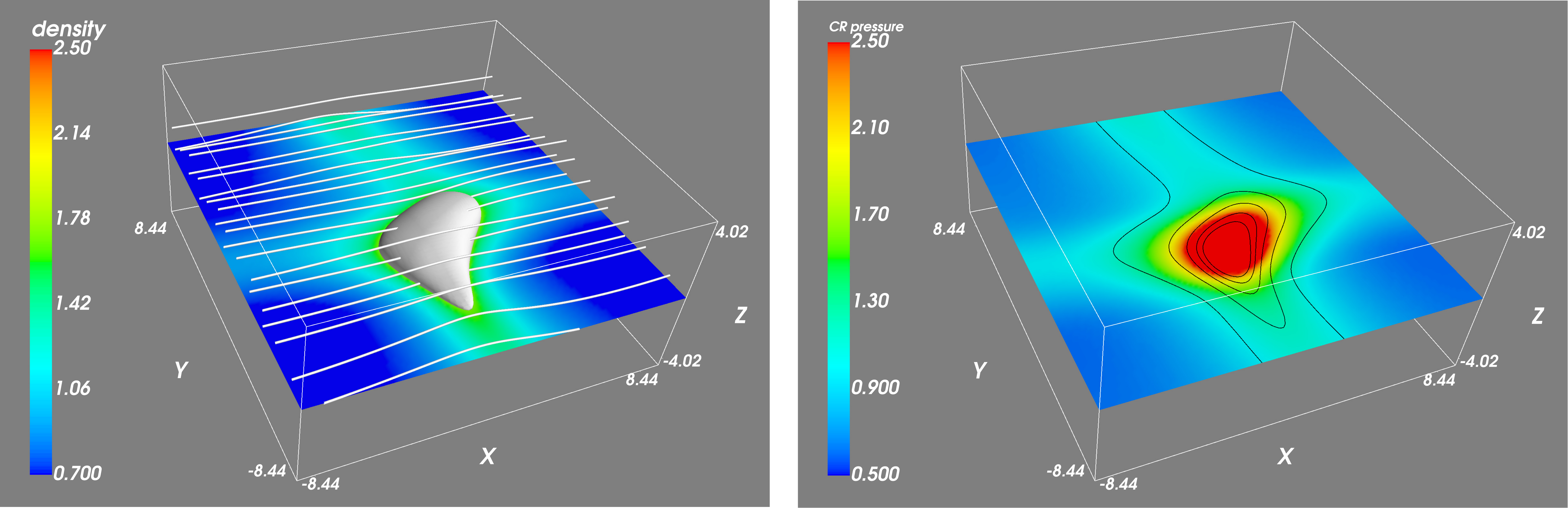}
\caption{Same as Figure~\ref{fig:model1} except for model~2
         ($\alpha=1.0$, $\beta=1.0$, $\kappa_\|=1.65$, $z_{\rm cor}=0.9$, $\Omega=0.0$) at $t\sim94\,{\rm Myr}$.}
\label{fig:model2}
\end{figure}

\begin{figure}
\plotone{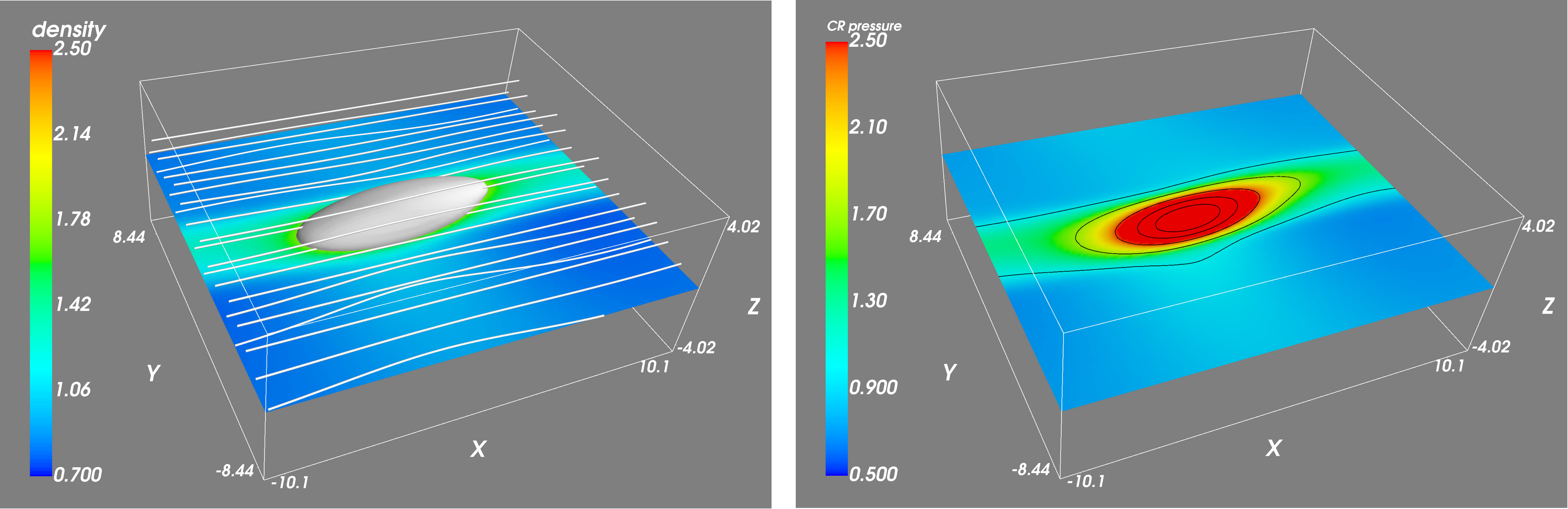}
\caption{Same as Figure~\ref{fig:model1} except for model~3
         ($\alpha=1.0$, $\beta=1.0$, $\kappa_\|=0.0$, $z_{\rm cor}=0.9$, $\Omega=0.0$) at $t\sim94\,{\rm Myr}$.}
\label{fig:model3}
\end{figure}

\begin{figure}
\plotone{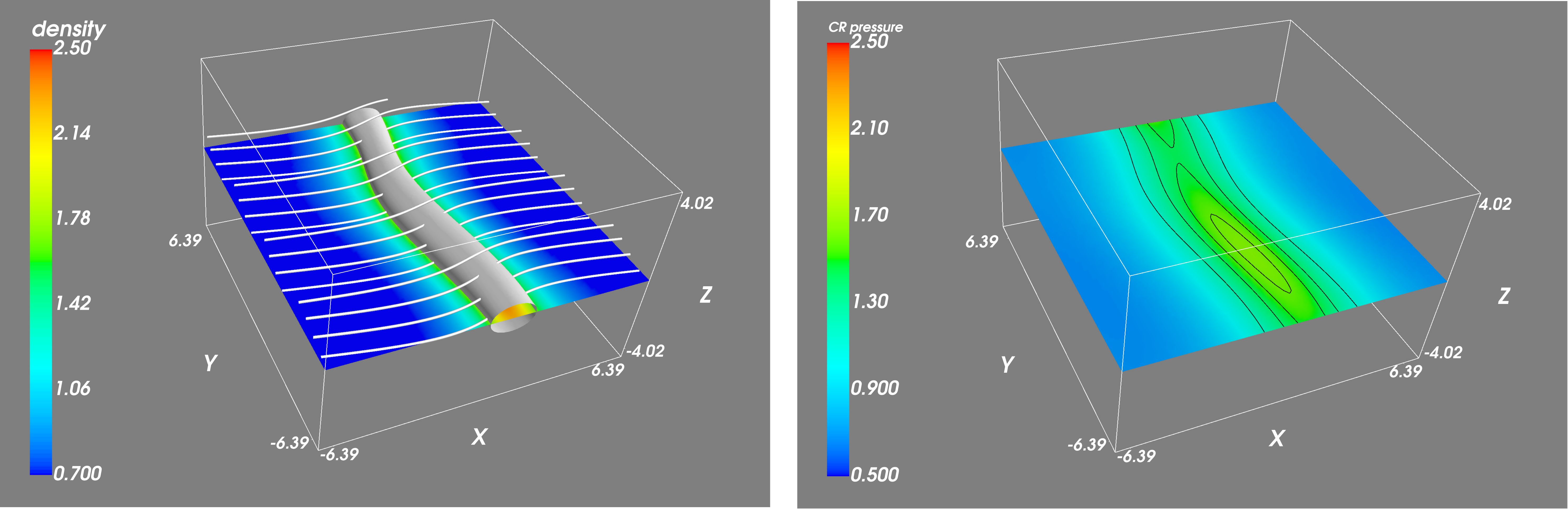}
\caption{Same as Figure~\ref{fig:model1} except for model~5
         ($\alpha=1.0$, $\beta=1.0$, $\kappa_\|=1.65$, $z_{\rm cor}=0.9$, $\Omega=0.3$) at $t\sim128\,{\rm Myr}$.}
\label{fig:model5}
\end{figure}

\begin{figure}
\plotone{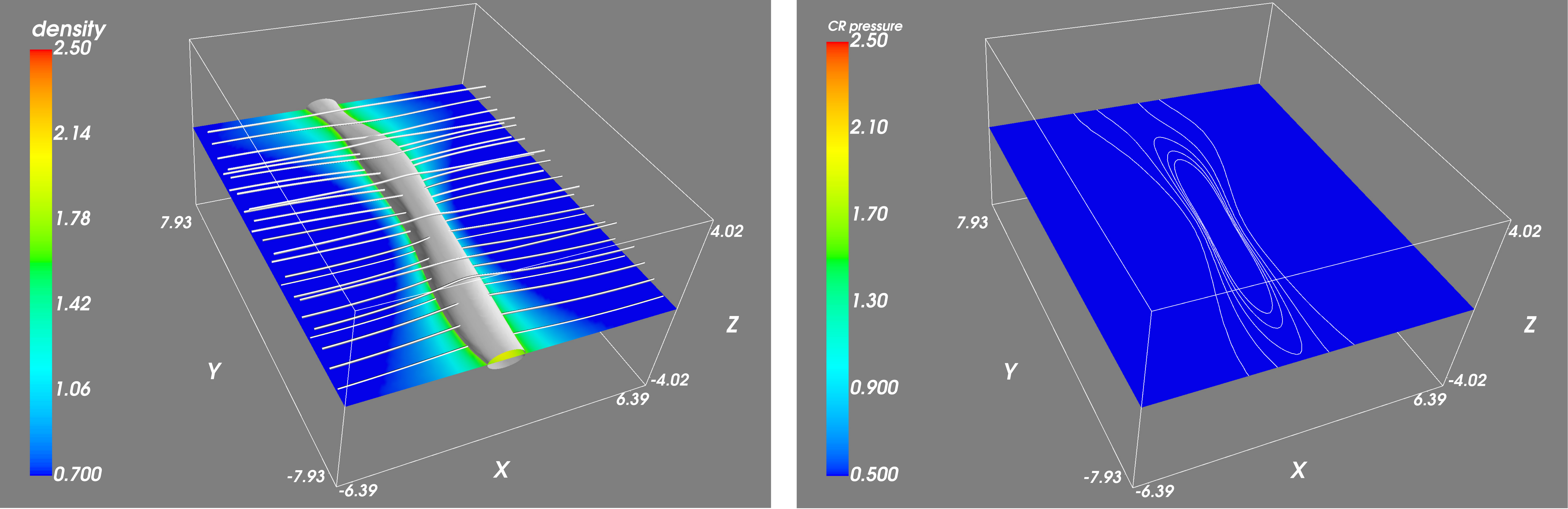}
\caption{Same as Figure~\ref{fig:model1} except for model~6
         ($\alpha=1.0$, $\beta=0.0$, $\kappa_\|=0.0$, $z_{\rm cor}=0.9$, $\Omega=0.0$) at $t\sim76\,{\rm Myr}$.}
\label{fig:model6}
\end{figure}

\begin{figure}
\plotone{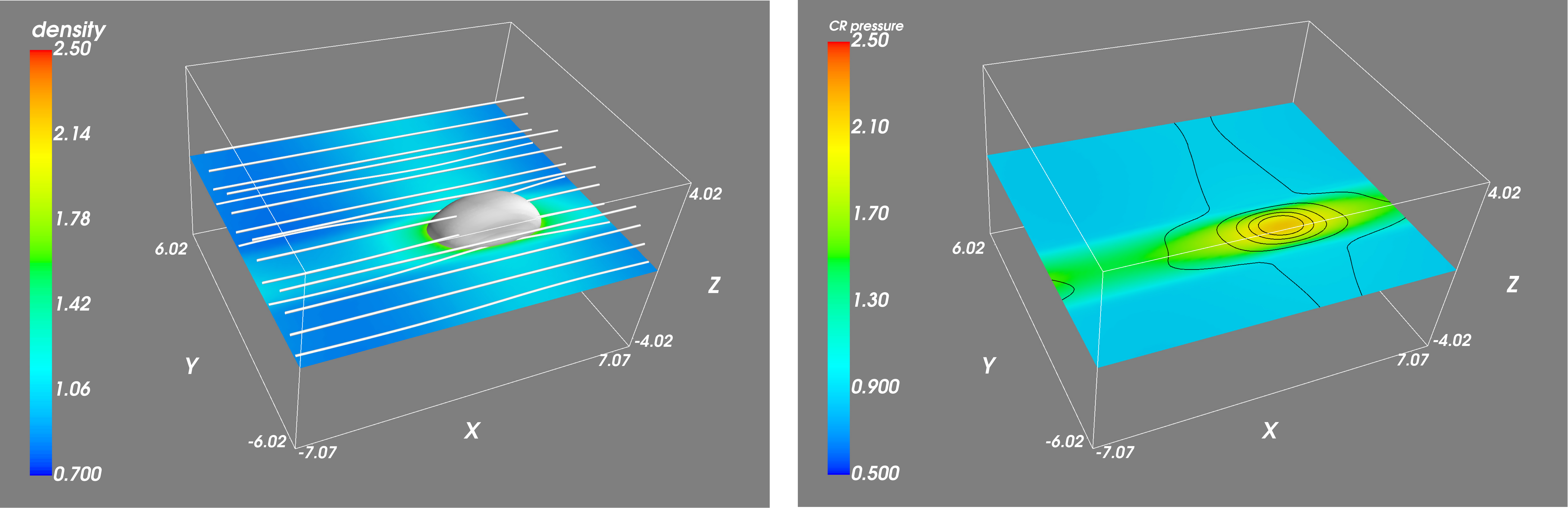}
\caption{Same as Figure~\ref{fig:model1} except for model~7
         ($\alpha=1.0$, $\beta=1.0$, $\kappa_\|=10.0$, $z_{\rm cor}=0.6$, $\Omega=0.0$) at $t\sim88\,{\rm Myr}$.}
\label{fig:model7}
\end{figure}

\begin{figure}
\plotone{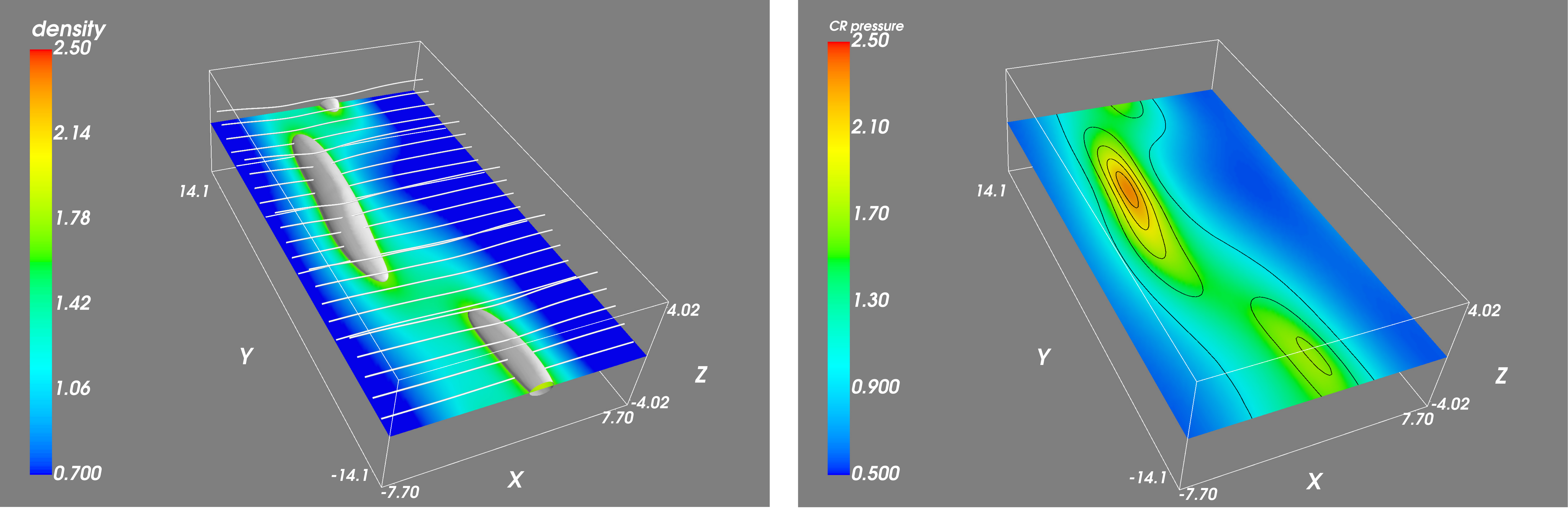}
\caption{Same as Figure~\ref{fig:model1} except for model~8
         ($\alpha=1.0$, $\beta=1.0$, $\kappa_\|=1.65$, $z_{\rm cor}=3.0$, $\Omega=0.0$) at $t\sim62\,{\rm Myr}$.}
\label{fig:model8}
\end{figure}

\begin{figure}
\plotone{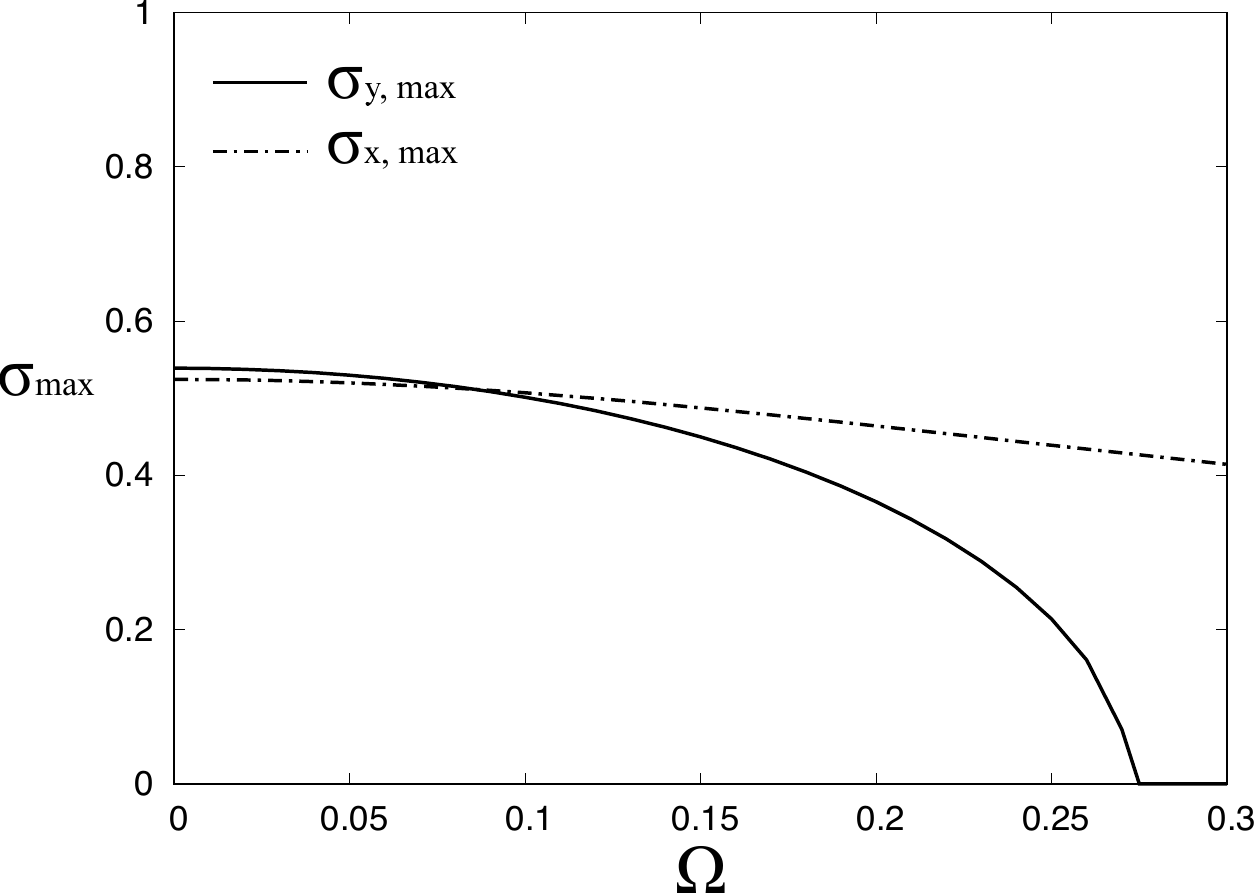}
\caption{Dependence of the maximum growth rate on angular velocity for
         $\alpha=1.0$, $\beta=1.0$, $\kappa_{\|}=1.65$ and $z_{\rm cor}=0.9$.
         $\sigma_{x\,{\rm max}}$ ($\sigma_{y\,{\rm max}}$) is the maximum growth rate of perturbations
         which does not depend on $y$ ($x$).}
\label{fig:sigma_omega}
\end{figure}

\begin{figure}
\plotone{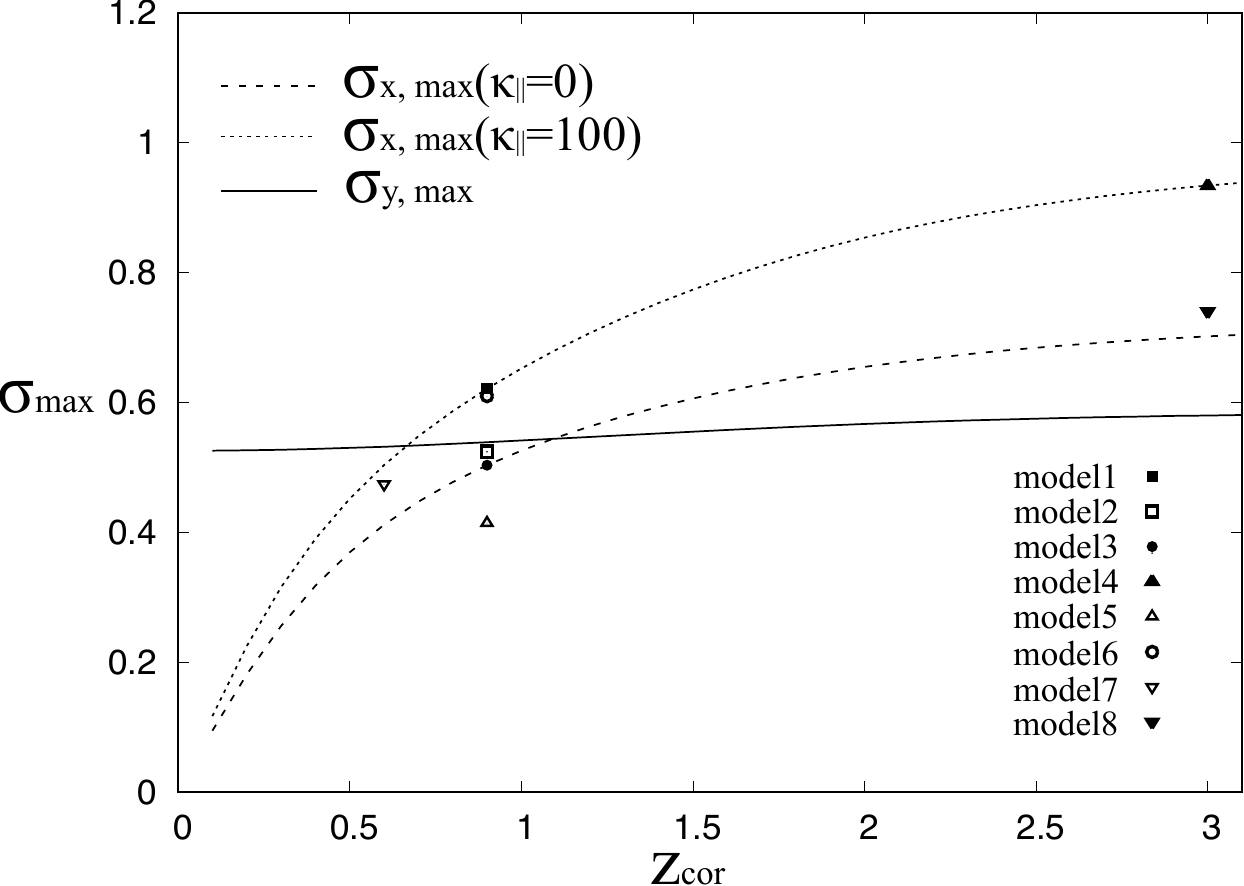}
\caption{Dependence of the maximum growth rates $\sigma_{x\,{\rm max}}$ and $\sigma_{y\,{\rm max}}$
         on the half thickness of the slab $z_{\rm cor}$ for the case $\alpha=1.0$, $\beta=1.0$ and $\Omega=0.0$.
         Note that $\sigma_{y\,{\rm max}}$ is independent of $\kappa_\|$.
         The $\sigma_{x\,{\rm max}}$ in each model is plotted for the convenience of comparing simulation results.}
\label{fig:gw_a1}
\end{figure}

\begin{figure}
\plotone{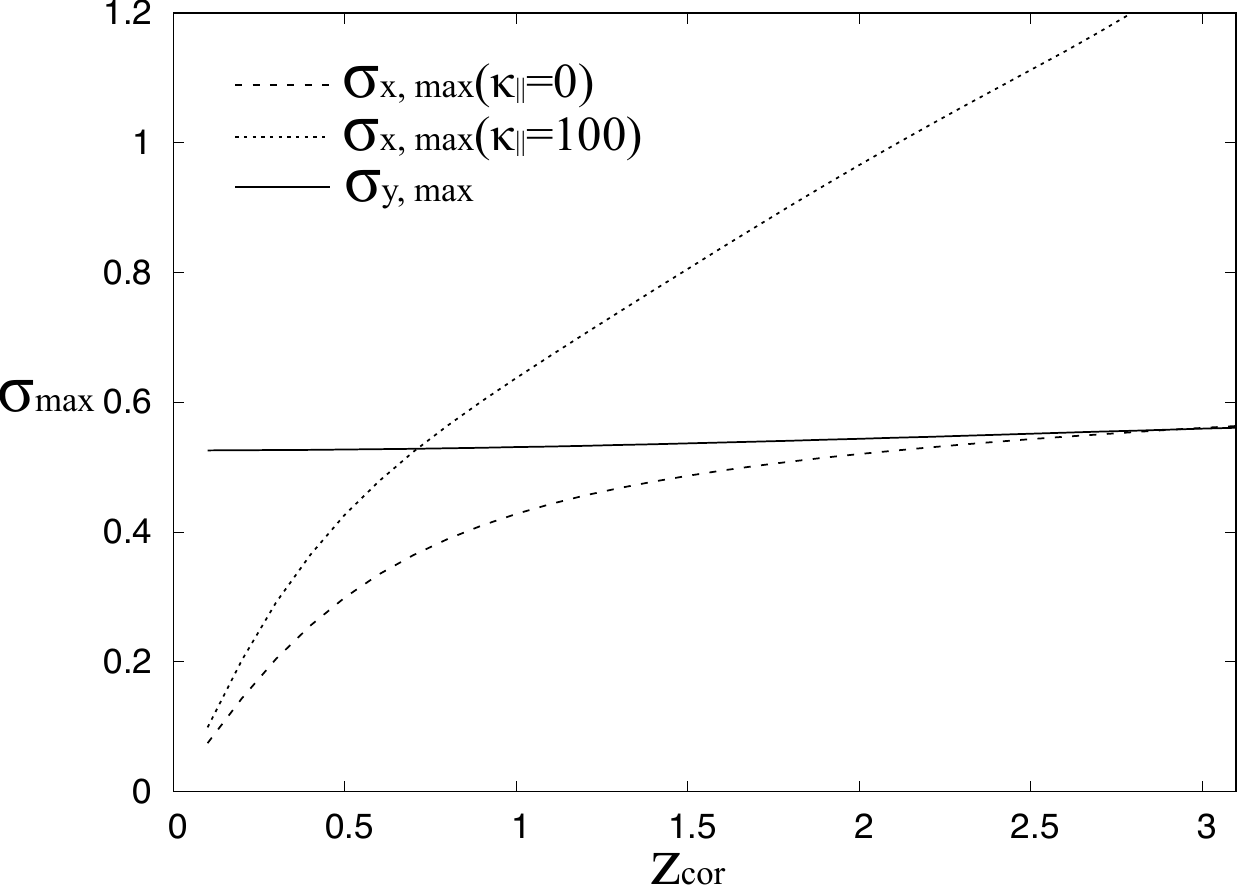}
\caption{Dependence of the maximum growth rates $\sigma_{x\,{\rm max}}$ and $\sigma_{y\,{\rm max}}$
         on the half thickness of the slab $z_{\rm cor}$ for the case $\alpha=1.0$, $\beta=10.0$ and $\Omega=0.0$.
         Note that $\sigma_{y\,{\rm max}}$ is independent of $\kappa_\|$.}
\label{fig:gw_a10}
\end{figure}

Generally speaking, the cold gas slab develops into filamentary structures,
but the direction of the filament with respect to the magnetic field
depends on the cosmic ray diffusion coefficient.
For the case of large diffusion coefficient the filaments form preferentially perpendicular to the magnetic field,
while for the case of small diffusion coefficient the filaments prefers to lie along the magnetic field.
For intermediate diffusion coefficient, clumps may form instead of filaments, and it will be impractical to
describe alignment or not.
These results agree well with the linear stability analysis by \citet{Kuwabara_2006}.
For illustration, we show the results of linear analysis in Figures~\ref{fig:gw_a1} and \ref{fig:gw_a10}.
The figures plot the maximum growth rates ($\sigma_{x\,{\rm max}}$, $\sigma_{y\,{\rm max}}$)
against the thickness of the gas slab $z_{\rm cor}$
(for the definition of $\sigma_{x\,{\rm max}}$ and $\sigma_{y\,{\rm max}}$,
see Section~\ref{subsec:results} or Figure~\ref{fig:sigma_omega}).
Note that $\sigma_{y\,{\rm max}}$ does not depend on $\kappa_\|$.
On the one hand, if $\sigma_{x\,{\rm max}}>\sigma_{y\,{\rm max}}$, then the instability variations grow faster in the $x$-direction
(the direction of magnetic field), thus the cold gas prefers to coalesce into filaments perpendicular to the magnetic field.
On the other hand, if $\sigma_{x\,{\rm max}}<\sigma_{y\,{\rm max}}$, the filaments form along the magnetic field.
In Figure~\ref{fig:gw_a1}, we observe that the thickness of the slab in the range $0.65\le z_{\rm cor}\le 1.1$ has an interesting feature:
$\sigma_{x\,{\rm max}}$ can be larger or smaller than $\sigma_{y\,{\rm max}}$ depending on the cosmic ray diffusion coefficient.
We call this the ``interchange zone''.
Therefore, the direction of the filaments formed from the Parker-Jeans instability depends on the diffusion coefficient.
As can be read from Figure~\ref{fig:gw_a1}, in the linear stage of model~1 ($\kappa_\|=100.0$),
$\sigma_{x\,{\rm max}}>\sigma_{y\,{\rm max}}$ which predicts that the filaments are perpendicular to the magnetic field.
This is exactly what the nonlinear stage of model~1 behaves (see Figure~\ref{fig:model1}).
Similarly, in the linear stage of model~3 ($\kappa_\|=0.0$), $\sigma_{x\,{\rm max}}<\sigma_{y\,{\rm max}}$,
and Figure~\ref{fig:model3} show exactly what is predicted: filaments form along the magnetic field.
In model~2, $\sigma_{x\,{\rm max}}\approx \sigma_{y\,{\rm max}}$, and the gas coalesces to form clumps.
Now, if we increase cosmic ray pressure from $\beta=1.0$ to 10.0,
the size of the ``interchange zone'' increases, see Figure~\ref{fig:gw_a10}.

\begin{figure}
\plotone{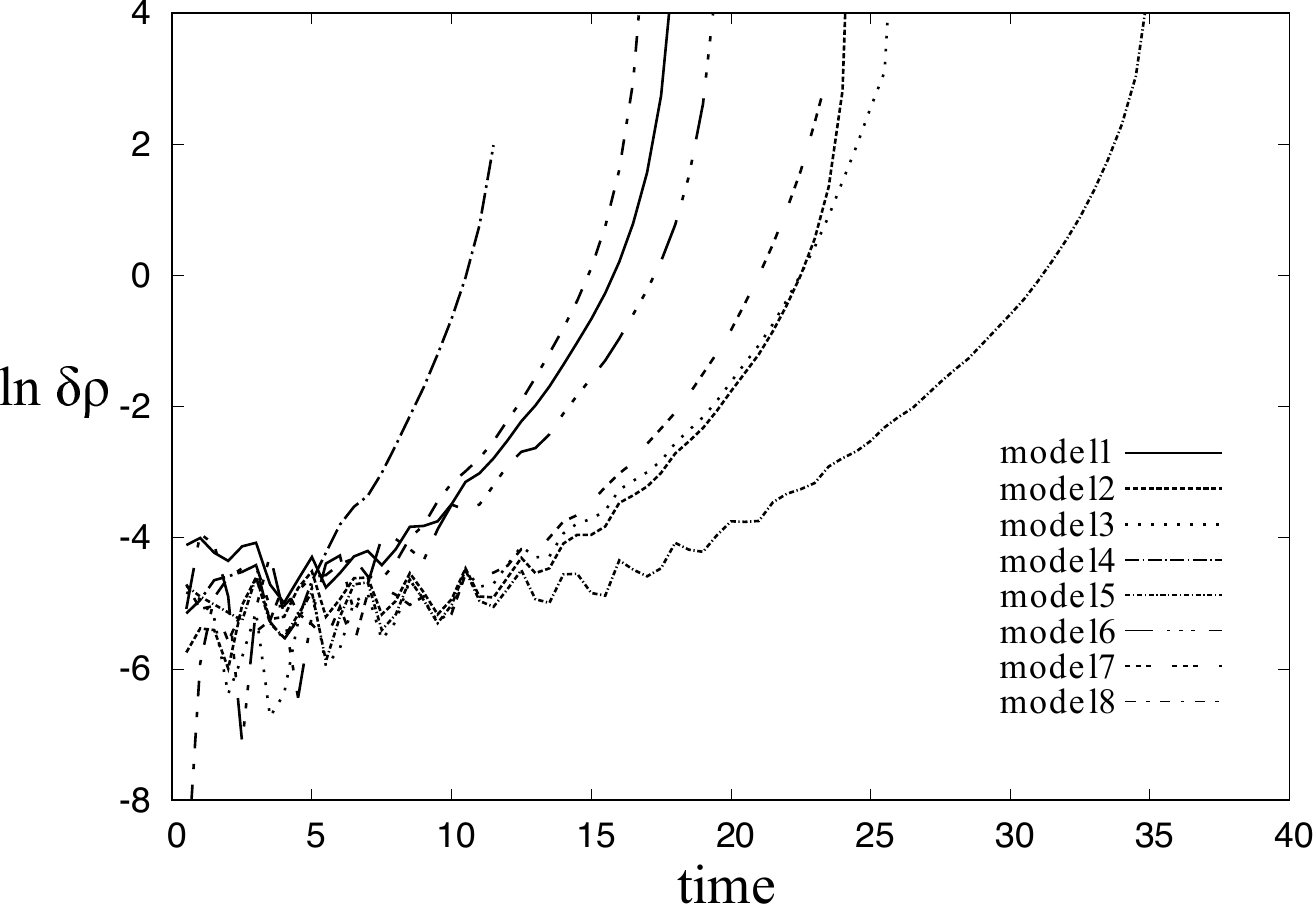}
\caption{Each curve show the time evolution of the perturbed gas density
         at the position where the density has its maximum value at the end of the simulation.}
\label{fig:time_den}
\end{figure}

Tracing the history of the evolution of the gas density at the position
where the density has its maximum value at the end of the simulation,
we learn that the growth rate depends on the cosmic ray diffusion coefficient, the thickness of the slab and rotation,
see Figure~\ref{fig:time_den}.
The growth rate increases as the diffusion coefficient increases or the thickness of the slab increases.
On the other hand, rotation suppresses the density growth rate and the suppression is different in
$\sigma_{x\,{\rm max}}$ and $\sigma_{y\,{\rm max}}$,
see Figure~\ref{fig:sigma_omega}.
$\sigma_{y\,{\rm max}}$ is strongly suppressed for large $\Omega$.

\begin{figure}
\plotone{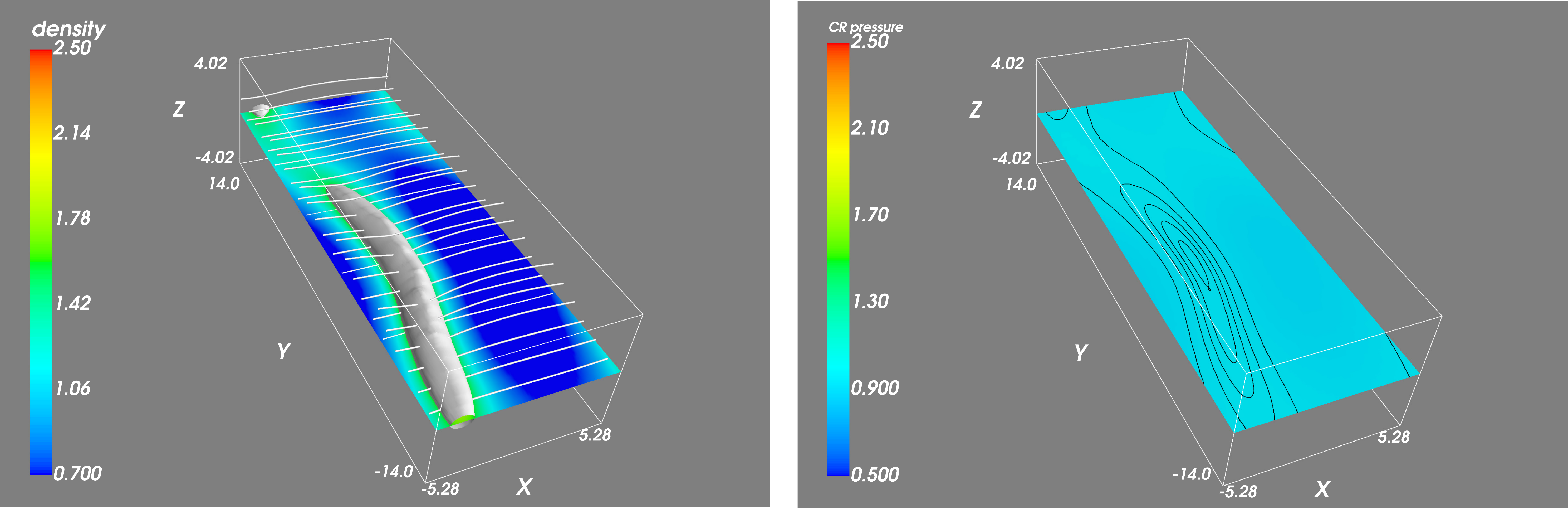}
\caption{Same as Figure~\ref{fig:model1} except for model~4
         ($\alpha=1.0$, $\beta=1.0$, $\kappa_\|=100.0$, $z_{\rm cor}=3.0$, $\Omega=0.0$) at $t\sim44\,{\rm Myr}$.}
\label{fig:model4}
\end{figure}

The influence of cosmic rays on the formation filaments or clumps from cold gas slab through
Parker-Jeans instability can be summarized in two parts: one on their action on Jeans instability and
the other on Parker instability.
As an addition fluid with significant pressure, cosmic rays help counter the self-gravity of the gas,
i.e., reduce of suppress Jeans instability.
However, if there is cosmic ray diffusion along the magnetic field, then the effect of cosmic ray
pressure on supporting the gas along the field lines reduces
while the effect has its full strength across the field lines
\citep[e.g., Appendix A of][]{Kuwabara_2006}.
Hence large diffusion coefficient along magnetic field exacerbates the tendency of cloud collapse along the field lines,
and the filaments preferentially form perpendicular to the magnetic field (e.g., Figures~\ref{fig:model1} \& \ref{fig:model4}).

As Parker instability develops, matter tends to slide down along magnetic field lines to the foot points and
coalesces to form clouds. However, in the case of small or zero diffusion coefficient, larger cosmic ray gradient is established
and impedes the matter motion towards the foot point \citep[][]{Kuwabara_2004}.
This facilitates the formation of filaments along the magnetic field (e.g., Figure~\ref{fig:model3}).

When the thickness of the gas slab is larger than the ``interchange zone'', Jeans instability dominates
and $\sigma_{x\,{\rm max}}$ is always larger than $\sigma_{y\,{\rm max}}$ (see Figure~\ref{fig:gw_a1} or \ref{fig:gw_a10}).
The filaments form perpendicular to the magnetic field.
On the other hand, when the thickness of the gas slab is smaller than the ``interchange zone'', it is conducive to Parker instability
and $\sigma_{x\,{\rm max}}$ is always smaller than $\sigma_{y\,{\rm max}}$ (see Figure~\ref{fig:gw_a1} or \ref{fig:gw_a10}).
In this case, the filaments prefers lying along the magnetic field.

Observations showed that the galactic magnetic fields are anchored at molecular clouds \citep[e.g.,][]{Han_2007,Li_2009,Li_2011}.
The existence of such ordered magnetic fields implies that the morphology of the magnetic field tends to be
preserved during the process of giant molecular cloud formation. 
Therefore, it is important to study the early stage of their formation and this work showed the effects of cosmic rays on this process.
Recently, bimodal distribution of orientation between cloud and magnetic field are observed in Gould Belt molecular clouds 
\citep[e.g.,][]{Li_2013,Planck_2016}.
\citet[][]{Soler_2013} confirmed it by the synthetic observation of the simulated turbulent molecular clouds.
In addition, \citet[][]{Soler_2017} showed that the direction change is an indication of compressive motions result from either gravitational collapse or
converging flows. 
In such phenomenon, the effect of cosmic ray diffusion may play an important role as shown in this work.

\clearpage

\begin{deluxetable}{cccccccc}
\tablenum{1}
\tablecaption{Models and parameters\label{tab:models}}
\tablewidth{0pt}
\tablehead{
\colhead{Model name} & \colhead{$\alpha$} & \colhead{$\beta$} & \colhead{$\kappa_{\|}$} &
\colhead{$z_{\rm cor}$} & \colhead{$\Omega$} &
\colhead{$\lambda_{x\,{\rm max}}$} & \colhead{$\lambda_{y\,{\rm max}}$}
}
\startdata
{model~1} & 1.0 & 1.0 & 100.0 & 0.9 & 0.0 & 12.67 & 16.89 \\
{model~2} & 1.0 & 1.0 & 1.65   & 0.9 & 0.0 & 16.89 & 16.89 \\
{model~3} & 1.0 & 1.0 & 0.0   & 0.9 & 0.0  & 20.14 & 16.89 \\
{model~4} & 1.0 & 1.0 & 100.0 & 3.0 & 0.0  & 10.56 & 28.05 \\
{model~5} & 1.0 & 1.0 & 1.65   & 0.9 & 0.3 & 12.77 & 0.0   \\
{model~6} & 1.0 & 0.0 & 0.0   & 0.9 & 0.0  & 12.77 & 15.87 \\
{model~7} & 1.0 & 1.0 & 10.0  & 0.6 & 0.0  & 14.15 & 12.04 \\
{model~8} & 1.0 & 1.0 & 1.65   & 3.0 & 0.0 & 15.40 & 28.30 \\
\enddata
\end{deluxetable}

\clearpage

\acknowledgments
CMK is supported in part by the Taiwan Ministry of Science and
Technology grant MOST 108-2112-M-008-006.

\bibliography{AAS23949R1}{}
\bibliographystyle{aasjournal}



\end{document}